\documentclass[12pt]{article}
\usepackage{amssymb}
\newcommand{\vek}[1]{\mbox{\bf #1}}
\begin{document}
\hspace*{\fill} LMU--TPW--97/14\\
\hspace*{\fill} hep-ph/9706278 \\[3ex]

\begin{center}
\Large\bf
Vector and Scalar Confinement in Gauge Theory with a Dilaton
\end{center}
\vspace{2ex}
\normalsize \rm
\begin{center}
{\bf Rainer Dick}\\[0.5ex] {\small\it
Sektion Physik der Universit\"at M\"unchen\\
Theresienstr.\ 37, 80333 M\"unchen,
Germany}
\end{center}

\vspace{5ex}
\noindent
{\bf Abstract}: 
In a recent letter it has been shown that gauge theory with a dilaton provides
linearly increasing gauge potentials from static or uniformly moving
pointlike colour sources. This ensures confinement in the framework
of no-pair equations.
Here I would like to point out that a dilaton coupling both
to the
gauge curvature term and to
fermion masses yields a linear potential with a scalar component
and a dominant vector contribution.

\newpage
\noindent
{\bf 1.\ }To understand confinement in gauge theories provides a
challenge for theoretical physics since more than two decades. 
Many approaches to the problem
rely on flux tube pictures,
with flux tubes emerging between
colour sources through gluon self-interactions or through
monopole condensation and the dual Meissner effect. 
Numerous investigations are dedicated to improve our understanding of this 
phenomenon and a collection of pioneering papers is \cite{pioneer}.
Meson spectroscopy indicates a linear growth of energy
with the distance between bound quarks and linear potentials
are under close investigation as phenomenological models for quark confinement
since many years, see \cite{early} for early references. Lattice calculations have 
confirmed this picture by showing
good agreement of a linear potential with a running coupling and nowadays are also
used to probe the spin and momentum dependent parts 
of quark interactions \cite{lattice}.
Important approaches to an improved theoretical understanding of confinement in 
gauge theories 
have been further developed over the years,
including semi-relativistic expansions of Wilson loops \cite{BP},
dual QCD \cite{BBZ}, and monopoles in  $N=2$
supersymmetric gauge theories \cite{SW}.

Recently it has also been pointed out that inclusion of a dilaton 
in gauge theories admits a solution
of the Coulomb problem with a linearly increasing gauge potential \cite{dil1}.
An important classical difference between linearly increasing gauge potentials
and $1/r$ potentials concerns the fact that the quark self energy is not ultraviolet
but infrared
divergent, whence the divergence could not be attributed to new physics at short
distances.
This indicates that increasing vector potentials
yield confinement. On the other hand, one would still like to confirm this
in a picture of gluon--dilaton mediated quark--quark interactions, where 
we assume that
at least
one of the quarks is heavy enough to make potential confinement a reasonable 
assumption.
It is well known that a linearly increasing 4--vector potential 
is sufficient to ensure confinement in the framework of
no-pair equations \cite{sucher}, and 
the assertion that the confining potential
is a Lorentz vector received wide acceptance in recent years.
On the other hand,
it is known since long that a linearly increasing timelike component of a 4--vector 
is not sufficient to ensure suppression of oscillatory solutions of the corresponding
Dirac equation without projection operators.
If one does not want to rely on no-pair
equations the gauge potential has to be supplemented
by a dominant scalar interaction to ensure 
confinement. I do not consider this as an attractive scenario, but I 
would like to point out that the dilaton can also
accomodate a scalar contribution to the confinement potential if in addition
to its coupling to $F^2$ it also couples to mass terms of quarks. The linearly
increasing scalar potential then arises as a consequence of the
logarithmic increase of the dilaton. However, we will see that the vector
contribution dominates. Although dominance of a scalar contribution
has been considered favourable for 
a while, dominance of the vector part saves us from a minor
complication: Scalar dominance in the interquark potential should imply that
all quark combinations are confined, with the phenomenologically
required combinations only being energetically preferred.
Nevertheless, by raising the quark masses above the coupling scale
of the dilaton we can identify a simple model where quark confinement
is realized even on the level of a naive Dirac equation.

Independently from the possibility to formulate models where confinement
can be treated analytically, there exists strong motivation to include
dilatonic degrees of freedom in gauge theory: 

Maybe the oldest motivation
comes from Kaluza--Klein theory, where the dilaton provides a measure
for the size of internal dimensions. From a more modern point of view,
the axion already provides
two independent reasons to include a dilaton in gauge theories:
If the gauge theory is supersymmetric the axion and the dilaton come together
in a chiral multiplet, and even without supersymmetry the axion has to come 
with a dilaton if duality symmetry is realized. The massless
spectrum of superstring theory provides yet another motivation to include
a dilaton, and recent developments reviewed e.g.\ in \cite{dual} 
encompass all these
reasons.

The possibility of a dilaton coupling to mass terms was not considered attractive
until a few years ago, since generically such a coupling is expected to conflict
constraints from the weak equivalence principle \cite{weak}. Indeed, for this
reason I assumed that the dilaton does not couple to low energy mass terms, the
justification for this assumption being that fundamental dilatons are expected 
to arise far
above any low energy scale where low energy mass terms arise e.g.\ through
spontaneous symmetry breaking.
On the other hand, Damour and Polyakov have pointed out that even massless
dilatons coupling to masses of other particles
may be in agreement with current observational constraints for
certain classes of coupling functions \cite{DP}, and this motivated me to
reconsider dilaton--mass couplings.\\[1ex]
{\bf 2.\ }In the present paper I would like to focus on a model
described by a Lagrange density
\begin{equation}\label{lagden}
{\cal L}=-\frac{1}{4}\exp\!\Big(\frac{\phi}{f_\phi}\Big)F_{\mu\nu}{}^j F^{\mu\nu}{}_j
-\frac{1}{2}\partial^\mu\phi\cdot\partial_\mu\phi
\end{equation}
\[
+\sum_{f=1}^{N_f}\overline{\psi}_f
(i\gamma^\mu\partial_\mu +g\gamma^\mu A_\mu{}^j X_j)\psi_f
-\exp\!\Big(-\xi\frac{\phi}{2f_\phi}\Big)\sum_{f=1}^{N_f}\overline{\psi}_f
m_f\psi_f,
\]
where $X_j$ denotes an $N_c$--dimensional representation
of su$(N_c)$. The particular form of the coupling functions of the dilaton
to gauge fields and fermions is not completely artificial: For
 $\xi=1$ it
corresponds exactly to what one would find in a Kaluza--Klein Ansatz
connecting Einstein--Yang--Mills theory in $D>4$ dimensions to
Einstein--Yang--Mills--dilaton theory in four dimensions,
where $\exp(2\phi/f_\phi)$ is the determinant of the internal metric.

The equations of motion are
\begin{equation}\label{eqmot1} 
\partial_\mu\Big(\exp\!\Big(\frac{\phi}{f_\phi}\Big)F^{\mu\nu}{}_i\Big)
+g\exp\!\Big(\frac{\phi}{f_\phi}\Big)A_\mu{}^j f_{ij}{}^k F^{\mu\nu}{}_k=
-g\overline{\psi}\gamma^\nu X_i\psi,
\end{equation}
\begin{equation}\label{eqmot2}
\partial^2\phi=\frac{1}{4f_\phi}\exp\!\Big(\frac{\phi}{f_\phi}\Big)F_{\mu\nu}{}^j F^{\mu\nu}{}_j
-\frac{\xi}{2f_\phi}\exp\!\Big(-\xi\frac{\phi}{2f_\phi}\Big)\overline{\psi}m\psi,
\end{equation}
\begin{equation}\label{eqmot3}
(i\gamma^\mu\partial_\mu +g\gamma^\mu A_\mu{}^j X_j)\psi
-\exp\!\Big(-\xi\frac{\phi}{2f_\phi}\Big)m\psi=0,
\end{equation}
Here and in the sequel flavour indices are suppressed.

We are interested in stationary colour distributions of the form
\[
j^\mu(x)=\varrho(\vek{r})\eta^\mu{}_0=\rho(\vek{r})C^i X_i\eta^\mu{}_0,
\]
where $C_i$ is the expectation value of the generator $X_i$.
Pointlike sources are of this type and the equations (\ref{eqmot1})
reduce to 
\[
\nabla\cdot\Big(\exp\!\Big(\frac{\phi}{f_\phi}\Big)\nabla\Phi\Big)=-\varrho,
\]
with $\Phi\equiv A^0$. For $\rho(\vek{r})=g\delta(\vek{r})$
we have 
\begin{equation}\label{sol1}
\exp\!\Big(\frac{\phi(r)}{f_\phi}\Big)\vek{E}_i(\vek{r})
=\exp\!\Big(\frac{\phi(r)}{f_\phi}\Big)E_i(r)\vek{e}_r=
\frac{gC_i}{4\pi r^2}\vek{e}_r
\end{equation}
and we have to determine the dilaton from
\begin{equation}\label{dil2}
\frac{d^2}{dr^2}\phi(r)+\frac{2}{r}\frac{d}{dr}\phi(r)=
-\frac{g^2}{64\pi^2f_\phi}\Big(1-\frac{1}{N_c}\Big)
\exp\!\Big(-\frac{\phi(r)}{f_\phi}\Big)\frac{1}{r^4}
-\frac{\xi}{2f_\phi}\exp\!\Big(-\xi\frac{\phi}{2f_\phi}\Big)m\delta(\vek{r}),
\end{equation}
where the property
\[
\sum_{i=1}^{N_c^2-1} C_i^2=\frac{N_c-1}{2N_c}
\]
of expectation values in colour space was used.
Using the abbreviation
\[
r_\phi=\frac{g}{8\pi f_\phi}\sqrt{\frac{1}{2}-\frac{1}{2N_c}}
\]
I would like to emphasize that both the regularized
Coulomb potential discovered in case $\xi=0$ in \cite{CT,dil1}
\begin{equation}\label{dilsol}
\phi(r)=2f_\phi
\ln\!\Big(\frac{r+r_\phi}{r}\Big),
\end{equation}
\begin{equation}\label{coulpot}
\Phi_i(r)=\frac{gC_i}{4\pi(r+r_\phi)},
\end{equation}
and the confining solution \cite{dil1}
\begin{equation}\label{dilsol2}
\phi(r)=
2f_\phi\ln\!\Big(\frac{r_\phi}{r}\Big),
\end{equation}
\begin{equation}\label{confpot}
\Phi_i(r)=
-\frac{gr}{4\pi r_\phi^2}C_i
\end{equation}
persist for arbitrary values $\xi\ge 0$.

The resulting scalar potentials seen by the fermions
in the two phases are
\[
S(r)=m\Big(\frac{r}{r+r_\phi}\Big)^\xi
\]
or
\[
S(r)=m\Big(\frac{r}{r_\phi}\Big)^\xi,
\]
respectively. \\[1ex]
{\bf 3.\ }To discuss the confining solution we assume that the potential is created by
a heavy pointlike source with an orientation $\zeta_s$ in colour space, i.e.\
\[
C_i=\zeta_s^+\cdot X_i\cdot\zeta_s.
\]
Restricting $\xi$ to the values $\xi=0$ (no mass coupling)
or  $\xi=1$ (Kaluza--Klein type coupling)
the net potential seen by an (anti-)quark of colour $\zeta_q$
in the rest frame of a heavy source
is then
\begin{equation}
V(r)=\Big[\pm\frac{g^2}{8\pi r_\phi}\Big(|\zeta_s^+\cdot\zeta_q|^2-\frac{1}{N_c}\Big)
+m\xi\Big]\frac{r}{r_\phi}.
\end{equation}
The vector part implies that a blue source attracts an anti-blue quark
in the sense that anti-quarks within an angle $\Theta_c=\arccos(\sqrt{1/N_c})$
from the positive or negative blue axis are attracted. Correspondingly a blue
source attracts quarks of different colour in the sense that their colour
orientation must lie outside the double cone defined by the angle $\Theta_c$.

The mass term, on the other hand, is always attractive, and we may ask
under which circumstances we could expect strict
confinement even in the framework
of a naive Dirac equation: The tree level cross section for creation of a dilaton
from head on collision of two gauge bosons with centre of mass energy $\sqrt{s}$
goes with $s/f_\phi^4$, and therefore we certainly expect
\[
m<gf_\phi\sqrt{2-\frac{2}{N_c}}
\]
no matter what definition of quark masses we would employ.
Therefore, the vector part will always dominate and we should not expect
scalar dominance in any realistic model. Nevertheless,  I would also like
to advertise the model with Kaluza--Klein type couplings and masses
exceeding the coupling scale $f_\phi$: 
These are models which provide confinement
in the simplest possible setting through the Dirac equation, and we may
learn something about thermodynamical aspects of confinement from 
these simple four-dimensional models.

Coming back to the vectorial part, we have seen that it is attractive
for quark--antiquark pairs of the same colour and for
quark pairs of different colour. Now suppose we have an ensemble
of quarks forming a plasma with a mean separation $r$.
Neglecting the scalar contribution, the energy gain $\epsilon_{b\bar{b}}$
in forming
a blue--anti-blue meson exceeds the energy gain $\epsilon_{br}$
in forming
a red--blue diquark by a factor $N_c-1$,
and therefore the gain in energy in forming $N_c$ mesons
equals the gain in energy in forming a nucleon and an anti-nucleon.
Hence, if the scalar contribution can be neglected and as 
long as quarks and anti-quarks
appear with equal densities the plasma has no preference energetically
to decay predominantly into one of the two channels. There nevertheless may
appear asymmetries for three reasons: Kinematically the formation of two-particle
bound states is favoured over the formation of $N_c$-particle bound states.
On the other hand, the scalar contribution would favour nucleons, since the
gain $2E_N$ in forming a nucleon and an anti-nucleon with the
mass couplings taken into account exceeds the gain $N_c E_{b\bar{b}}$
in forming $N_c$
mesons by $2E_N=N_c E_{b\bar{b}}+N_c(N_c-2)\Delta$. Here $\Delta$ is the contribution
from the scalar term to the binding energy of a meson or diquark. Finally, 
only nucleons would form
after the anti-quarks have been used up.

\end{document}